# Mechanical properties of borophene films: A reactive molecular dynamics investigation

Minh-Quy Le[3,\*], Bohayra Mortazavi[4,\*], Timon Rabczuk[1,2,4,\*]

[1]Division of Computational Mechanics, Ton Duc Thang University, Ho Chi Minh City, Vietnam

[2]Faculty of Civil Engineering, Ton Duc Thang University, Ho Chi Minh City, Vietnam

[3]Department of Mechanics of Materials and Structures, School of Mechanical Engineering, Hanoi University of Science and Technology, No. 1, Dai Co Viet Road, Hanoi, Vietnam

[4]Institute of Structural Mechanics, Bauhaus-Universität Weimar, Marienstr. 15, D-99423 Weimar, Germany

**Abstract**- Most recent experimental advances could provide ways for the fabrication of several atomic thick and planar forms of boron atoms. For the first time we explore mechanical properties of 5 types of boron films with various vacancy ratios ranging from 0.1 to 0.15, using molecular dynamics simulations with ReaxFF force field. It is found that the Young's modulus and tensile strength decrease with increasing the temperature. We found that boron sheets exhibit anisotropic mechanical response due to different arrangement of atoms along armchair and zigzag directions. At room temperature, two-dimensional Young's modulus and fracture stress of these 5 sheets appear in the range 63-136 N/m and 12-19 N/m, respectively. In addition, strains at tensile strength are in the ranges of 9-14%, 11-19%, and 10-16% at 1, 300, and 600 K, respectively. This investigation not only reveals remarkable stiffness of 2D boron but also establish relations between the mechanical properties of boron sheets to loading direction, temperature and atomic structures.

\* Corresponding authors: quy.leminh@hust.edu.vn;
bohayra.mortazavi@gmail.com; timon.rabczuk@uni-weimar.de

# 1. Introduction

Boron has diverse forms of allotropes [1, 2], including 0D clusters, 1D nanotubes and nanowires, and various 2D sheets. Boron analogs of planar carbon allotropes such as graphene have been recently theoretically predicted [3-7]. These 2D boron sheets contain of hexagonal and triangular lattices and they present metallic electronic properties. Exciting experimental advances have just taken place with respect to the synthesis of two-dimensional (2D) boron films so called borophene [8, 9]. Boron sheets with the ratio $\eta$ between the number of hexagon holes (or number of missing atoms or vacancies from the original triangular sheet) and the number of atoms in the original triangular sheet ranging from 0.1-0.15 are the most stable ones and planar [3, 4, 6]. Further, Zhang et al. [3] showed that structures of boron monolayers depend on metallic substrates on which the boron layers are formed. Among all boron sheets predicted theoretically, triangular boron sheet ($\eta$=0) and boron sheet with $\eta$=1/5 were recently grown on Ag substrates [8, 9], $\gamma$-boron sheet ($\eta$=1/6) was synthesized on copper foils [10] and on Ag substrate [8]. So far, the studies on mechanical and physical properties of boron sheets are quite limited. Peng et al. [11] investigated mechanical properties of $\alpha$-boron sheets ($\eta$=1/9) by density functional theory (DFT) calculations. Electronic, magnetic, and optical properties of triangular boron sheets were recently investigated by DFT calculations [12, 13]. 2D boron sheets were theoretically predicted to exhibit a superconductivity [14].

Motivated by the most recent experimental advances in the fabrication of borophene films [3, 8, 9, 15], various properties of 2D boron sheets should be studied in detail to provide better viewpoint concerning their application prospects. Because of the complexity, uncertainties and costs of experimental characterization techniques for the evaluation of properties of atomic thick materials, theoretical and modeling approaches can be considered as promising approaches [16, 17]. Among different material properties, mechanical properties not only play crucial role for the strength and durability of the performance of a material in service but also provide a data base to help designing nano-devices. In the present study, we aimed to provide a general vision regarding the mechanical properties of several borophene films by performing extensive classical but reactive molecular dynamics (MD) simulations. It should be noted that based on the experimental findings, although the borophene sheets were grown on silver they interact weakly with their substrate [**8, 9**]. This implies that experimental evidence of free-standing borophene is highly probable in near future. Therefore, in this study we only considered the mechanical responses of free-standing and single-layer borophene films. Nevertheless, possible effect of interactions with substrate on the mechanical properties of boron sheets should be studied in separate investigations. We studied the mechanical properties of various free-standing boron monolayers at 1 K, 300 K, and 600 K. Young's modulus, fracture stress and strain are evaluated and the fracture mechanism is also analyzed in detail.

## 2. Method

ReaxFF potential has been used to model B-B interactions [**18**]. MD simulations were carried out using LAMMPS code [**19**]. Verlet time-integration algorithm is used [**20**]. Periodic boundary condition is applied in the planar directions (*x*- and *y*-directions in this work) to remove the finite-length effect. This way, we studied borophene films and not the 2D boron nanoribbon with free edges. We studied the mechanical properties of single-layer boron sheets, hence, we used a 20 Å vacuum along the thickness of the films (z-direction) by defining a fixed simulation box size. MD simulations were carried out at 1, 300, and 600 K. The structures were relaxed to zero stress using the Nosé-Hoover barostat and thermostat (NPT) method for 1.25 ps. Uniaxial tensions were carried out in the armchair and zigzag directions (notations are indicated in Fig. 1) by applying a constant engineering strain rate of $2.5 \times 10^8$ s$^{-1}$ in the tensile direction. Based on the applied engineering strain rate and the system simulation box size, the velocity of the loading was calculated. Because we are dealing with single-layer boron sheets, the stress in the sheet thickness is zero (*z*-direction). Therefore, in order to guarantee uniaxial stress conditions, zero stress condition in the edge parallel to the tensile direction was achieved by altering the size of simulation box in the direction perpendicular to the tensile direction using the NPT method.

5 boron sheets with $\eta=1/9$ ($\alpha$-boron sheet), 1/8 (type A), 1/8 (type B), 2/15, 4/27, are here investigated. These 5 sheets with $\eta=0.1$-$0.15$ are planar and most stable ones [3, 4, 6]. Table 1 summaries the geometric dimension of boron sheets at relaxation at 0.001 K.

Macroscopic stress tensor is estimated by using the virial theorem [21, 22]:

$$\boldsymbol{\sigma} = \frac{1}{V}\sum_{a\in V}\left[-m^{(a)}\mathbf{v}^{(a)}\otimes\mathbf{v}^{(a)} + \frac{1}{2}\sum_{a\neq b}\mathbf{r}^{(ab)}\otimes\mathbf{f}^{(ab)}\right]. \qquad (1)$$

$m^{(a)}$ and $\mathbf{v}^{(a)}$ are the mass and velocity vector of atom $a$, respectively. The symbol $\otimes$ denotes the tensor product of two vectors. $\mathbf{r}^{(a)}$ denotes the position of atom $a$. $\mathbf{r}^{(ab)} = \mathbf{r}^{(b)} - \mathbf{r}^{(a)}$ is the distance vector between the atoms $a$ and $b$. $\mathbf{f}^{(ab)}$ is the interatomic force applied on atom $a$ by atom $b$. $V$ is the volume of the structure. For boron sheet, $V = A.t$, where $A$ is the surface area of the sheet, and $t$ is the sheet's thickness.

Young's modulus $Y$ is determined from the first derivative of the stress-strain curve at $\varepsilon = 0$. Due to an ambiguous value of the sheet's thickness, we use $Yt$ and $\sigma t$ for 2D Young's modulus (or in plane-stiffness) and 2D stress (or in-plane stress), respectively. However, in the remainder of the paper we will use Young's modulus $Yt$ and axial stress $\sigma t$ for short.

## 3. Results and discussion

Stress-strain curves under uniaxial tension in the armchair and zigzag directions are plotted in Fig. 2-3 for various sheets at 1, 300, and 600 K. At 1 and 300 K, At 1 and 300 K, fracture occurs suddenly and a drop in the stress–strain curve is observed for all sheets as indicated in Figs. 2 and 3. In these cases of brittle fracture, maximal axial stress (or tensile strength) and strain at maximal stress refer to fracture stress and fracture strain, respectively. At 600 K, under tension in the armchair direction, the sheets with $\eta=1/9$, 1/8 (A), 2/15, 4/27 do not exhibit brittle fractures as shown in Fig. 3a, 3b, 3d and 3e. Failure does not occur immediately after the point of maximal stress. Tension of these sheets prolong significantly after the point of maximal stress. Strains at maximal stress are 10.3 %, 10.7 % and 15.6 %; while final fracture took place at axial strain of about 19.4 %, 32 %, and 19.4 % for the sheets with $\eta$=1/9, 1/8 (A), 4/27 under tension in the armchair direction at 600 K, respectively.

Young's modulus of $\alpha$-boron sheet ($\eta$=1/9) is about 249, 128, and 109 N/m (in the zigzag direction); and 284, 136, and 115 N/m (in the armchair one) at 1, 300, and 600 K respectively. Our estimated Young's modulus of $\alpha$-boron sheet at 1 K is comparable with that of 224.6 N/m from DFT calculations [**11**]. Using other ReaxFF force field parameters [**23**], Young's modulus of $\alpha$-boron sheet at 1 K is estimated at about 167 and 144 N/m in the zigzag and armchair direction, respectively. It is checked that available Tersoff potentials [**24, 25**] fails to relax complicated atomic structures of $\alpha$-boron sheet.

Table 2 summaries Young's modulus, tensile strength, strain at tensile strength, and toughness of various boron sheets. The toughness is the area under the stress-strain curve till tensile strength. Effects of temperatures on Young's modulus, tensile strength, strain at tensile strength, and toughness are shown in Fig. 4-7 for the 5 sheets with $\eta$=0.1-0.15. In general, Young's modulus and tensile strength decrease with an increase of temperature. Decrease of Young's modulus and tensile strength from temperature of 1 to 300 K is larger than that from temperature of 300 to 600 K as shown in Fig. 4. When increasing temperature from 1 to 300 K, Young's modulus reduces about 47-67% and 49-57%, in the zigzag and armchair directions, respectively. A reduction in Young's modulus of 5-21% in the zigzag direction and 8-29% in the armchair one are found for a further increase of temperature from 300 to 600 K. At 300 K, Young's modulus is about 128, 123, 63, 119, and 95 N/m in the zigzag direction; and 136, 126, 126, 116, and 96 N/m in the armchair direction for the sheets with $\eta$=1/9, 1/8 (A), 1/ 8 (B), 2/15, and 4/27, respectively.

When increasing temperature from 1 to 300 K, tensile strength reduces about 10-62% and 41-63%, in the zigzag and armchair directions, respectively. Tensile strength decreases about 17-39% and 26-35% when temperature rises from 300 to 600K. At 300 K, tensile strength is about 14.7, 12.3, 12.0, 19.4, and 14.0 N/m in the zigzag direction; and 14.1, 15.2, 18.4, 15.0, and 13.1 N/m in the armchair direction for the sheets with $\eta$=1/9, 1/8 (A), 1/8 (B), 2/15, and 4/27, respectively.

In both zigzag and armchair directions, strains at tensile strength are about 9-14%, 11-19%, and 10-16% at 1, 300, and 600 K, respectively. Strains at tensile strength of the 5 sheets with $\eta$=0.1-0.15 increase when increasing temperature from 1 to 300 K.

Under tension in the zigzag direction of the sheet with $\eta$=1/8 (A), strains at tensile strength are slightly affected by the temperature (about 10.5, 11.0, and 11.2% at 1, 300, 600 K, respectively). Slight increases in strains at tensile strength under tension in the armchair direction are found for the sheets with $\eta$=2/15 (14.3% at 300 K and 15.1% at 600 K) and $\eta$=4/27 (15.0% at 300 K and 15.6% at 600 K) when increasing temperature from 300 to 600 K. Strains at tensile strength of other cases increase with an increase of temperature from 1 to 300 K, then decrease for a further increase of temperature from 300 to 600 K as shown in Fig. 6. Hence, strains at tensile strength are highest at 300 K (among 3 regimes of temperature) including: under tension in the armchair direction of the sheets with: $\eta$=1/9 (11.8%), $\eta$=1/8 (B) (16.5%), $\eta$=2/15 (14.3%), $\eta$=4/27 (15.0%); and under tension in the zigzag direction of the sheets with: $\eta$=1/9 (15.4%), $\eta$=1/8 (B) (17.3%), $\eta$=2/15 (18.6%), $\eta$=4/27 (18.0%). It should be noted that mechanical properties of graphene decreases slightly when increasing temperature in the range 100-600 K [**26**].

The toughness of boron sheets decreases in general with an increase of temperature as shown in Fig. 7. The toughness of 5 studied boron sheets appears in the range: 1.2-2.0; 0.65-1.8; and 0.6-1.0 J.m$^{-2}$ at 1, 300, and 600 K, respectively. At room temperature, among 5 sheets the sheet with $\eta$=2/15 exhibits the highest toughness, about 1.8 J.m$^{-2}$ and 1.0 J.m$^{-2}$ in the zigzag and armchair direction, respectively.

A comparison of mechanical properties of 2 boron sheets with $\eta$=1/8 (type A) and $\eta$=1/8 (type B) is shown in Fig. 8. Atomic structure of these 2 sheets with $\eta$=1/8 is schematically illustrated in Fig. 1c and 1d. Under tension in the zigzag direction, the sheet with $\eta=1/8$ (type A) exhibits higher Young's modulus, higher tensile strength, and lower strain at tensile strength than those of sheet with $\eta=1/8$ (type B). Under tension in the armchair direction, the sheet with $\eta=1/8$ type A exhibits lower tensile strength, and lower strain at tensile strength than those of sheet with $\eta=1/8$ type B, while their Young's moduli are fairly different.

Effects of vacancy ratio $\eta$ on Young's modulus, tensile strength, strain at tensile strength, and toughness are indicated in Fig. 9. Fig. 9 includes the sheet with $\eta=1/8$ type A and excludes the sheet with $\eta=1/8$ type B. At a given $\eta$, boron sheet exhibits anisotropic because its mechanical properties are different in the armchair and zigzag directions as clearly seen in Fig. 9. In general, Young's modulus decreases with an increase of the vacancy ratio $\eta$ as shown in Fig. 9a and Table 2. At 1 K, tensile strength decreases with an increase of the vacancy ratio $\eta$. However, tensile strengths are fairly different at 600 K for 4 sheets with $\eta=1/9$, 1/8 (A), 2/15 and 4/27. It should be emphasized that even a single vacancy could reduce significantly the Young's modulus, fracture stress and fracture strain in graphene [**27, 28**], boron-nitride [**29**], and silicene [**30**].

Strain at tensile strength increases with an increase of the vacancy ratio $\eta$ at 1 and 600 K as shown in Fig. 9c. It is noted that an increase of fracture strain was also found in graphene when the defect ratio falls in the range from 7-12% [**31**]. Fig. 9d shows effects of vacancy on the toughness of boron sheets. At 600 K, the toughness increases with an increase of the vacancy ratio $\eta$ under tension in both armchair and zigzag directions. At 300 K this tendency is clearly found only under tension in the armchair direction.

Fig. 10-15 shows snapshots of boron sheets under certain stages of tension. During tension, hexagons enlarge. Fracture mechanism is in general the coalescence of local failures at hexagon holes.

# 4. Conclusions

Extensive reactive modelcular dunamics simulations were carried out to investigate mechanial properties of 5 boron sheets with various vacancy concentrations ranging from 0.1 to 0.15 at 3 temperatures 1, 300, and 600 K. Our findings can be summarized as follows:

as below:

- Several reactive and non-reactive force fields were used to model the 2D boron using the classical molecular dynamics simulation. We found that latest ReaxFF force field developed by Pai et al. [**18**] provide the most accurate results with respect to the first principles predictions available in the literature.

- Young's modulus and tensile strength decrease with an increase of temperature. Decrease of Young's modulus and tensile strength from temperature of 1 to 300 K is larger than that from temperature of 300 to 600 K. In both zigzag and armchair directions, strains at tensile strength are about 9-14%, 11-19%, and 10-16% at 1, 300, and 600 K, respectively. All studied boron sheet exhibit anisotropic because their mechanical properties are different in the armchair and zigzag directions.

- At 300 K, Young's modulus is about 128, 123, 63, 119, and 95 N/m in the zigzag direction; and 136, 126, 126, 116, and 96 N/m in the armchair direction for the sheets with $\eta$=1/9, 1/8 (A), 1/8 (B), 2/15, and 4/27, respectively. At 300 K, tensile strength is about 14.7, 12.3, 12.0, 19.4, and 14.0 N/m in the zigzag direction; and 14.1, 15.2, 18.4, 15.0, and 13.1 N/m in the armchair direction for the sheets with $\eta$=1/9, 1/8 (A), 1/8 (B), 2/15, and 4/27, respectively.

- In general, Young's modulus decreases with an increase of the vacancy ratio $\eta$. Strain at tensile strength increases with an increase of the vacancy ratio $\eta$ at 1 and 600 K. At 1 K, tensile strength decreases with an increase of the vacancy ratio $\eta$. However, tensile strengths are fairly different at 600 K for 4 sheets with $\eta$=1/9, 1/8 (A), 2/15 and 4/27.

- Fracture mechanism is in general the coalescence of local failures at hexagon holes. At 1 and 300 K, all sheets exhibit brittle fracture. At 600 K, most of cases are also brittle fracture. However, brittle fracture was not observed under tension in the armchair direction at 600 K of the sheets with $\eta$=1/9, 1/8 (A), 4/27. These 3 sheets prolong significantly after the point of tensile strength.

• In general, toughness of boron sheets decreases with an increase of temperature. The toughness of 5 studied boron sheets appears in the range: 1.2-2.0; 0.65-1.8; and 0.6-1.0 J.m$^{-2}$ at 1, 300, and 600 K, respectively. At 600 K, the toughness increases with an increase of the vacancy ratio $\eta$ under tension in both armchair and zigzag directions. At 300 K this tendency is only clearly found under tension in the armchair direction.

## Acknowledgements

MQL was supported by the Alexander Von Humboldt Foundation under the renewed research program. BM and TR greatly acknowledge the financial support by European Research Council for COMBAT project (Grant number 615132).

Table 1 Geometric dimension of boron sheets at 0.001 K used in MD simulations

| $\eta$ (sheet type) | Length in zigzag direction, Å | Length in armchair direction, Å | Number of atoms |
| --- | --- | --- | --- |
| 1/9 ($\alpha$-boron sheet) | 143.86 | 149.03 | 5600 |
| 1/8 (type A) | 136.91 | 137.12 | 4830 |
| 1/8 (type B) | 136.89 | 137.16 | 4830 |
| 2/15 | 136.90 | 137.16 | 4784 |
| 4/27 | 144.19 | 143.53 | 5152 |

Table 2 Mechanical properties of boron sheets (*: DFT by Peng et al. [11]))

| $\eta$ (sheet type) | Young's modulus $Y_s$, N/m | Tensile strength, N/m | Strain at tensile strength, % | Toughness, J.m$^{-2}$ |
|---|---|---|---|---|
| 1/9 ($\alpha$ sheet), 1K | 249 (zigzag) | 33.1 (zigzag) | 8.66 (zigzag) | 1.27 (zigzag) |

| | | | | | |
|---|---|---|---|---|---|
| | | 284 (armchair) | 38.1 (armchair) | 10.0 (armchair) | 1.86 (armchair) |
| | | 224.6* | | | |
| | 300K | 128 (zigzag) | 14.7 (zigzag) | 15.41 (zigzag) | 1.21 (zigzag) |
| | | 136 (armchair) | 14.1 (armchair) | 11.84 (armchair) | 0.75 (armchair) |
| | 600K | 109 (zigzag) | 11.2 (zigzag) | 9.9 (zigzag) | 0.65 (zigzag) |
| | | 115 (armchair) | 10.4 (armchair) | 10.3 (armchair) | 0.51 (armchair) |
| 1/8 (type A), 1K | | 288 (zigzag) | 32.3 (zigzag) | 10.5 (zigzag) | 1.69 (zigzag) |
| | | 284 (armchair) | 29.6 (armchair) | 9.18 (armchair) | 1.31 (armchair) |
| | 300K | 123 (zigzag) | 12.3 (zigzag) | 11.0 (zigzag) | 0.65 (zigzag) |
| | | 126 (armchair) | 15.2 (armchair) | 14.16 (armchair) | 1.00 (armchair) |
| | 600K | 97 (zigzag) | 10.2 (zigzag) | 11.2 (zigzag) | 0.59 (zigzag) |
| | | 114 (armchair) | 10.8 (armchair) | 10.7 (armchair) | 0.62 (armchair) |
| 1/8 (type B), 1K | | 189 (zigzag) | 13.4 (zigzag) | 13.7 (zigzag) | 1.16 (zigzag) |
| | | 290 (armchair) | 36.9 (armchair) | 11.7 (armchair) | 1.69 (armchair) |
| | 300K | 63 (zigzag) | 12.0 (zigzag) | 17.3 (zigzag) | 0.98 (zigzag) |
| | | 126 (armchair) | 18.4 (armchair) | 16.5 (armchair) | 1.62 (armchair) |
| | 600K | 60 (zigzag) | 9.1 (zigzag) | 14.1 (zigzag) | 0.66 (zigzag) |
| | | 99 (armchair) | 13.1 (armchair) | 13.75 (armchair) | 0.97 (armchair) |
| 2/15, 1K | | 280 (zigzag) | 33.4 (zigzag) | 11.8 (zigzag) | 2.07 (zigzag) |
| | | 228 (armchair) | 25.5 (armchair) | 9.85 (armchair) | 1.19 (armchair) |
| | 300K | 119 (zigzag) | 19.4 (zigzag) | 18.6 (zigzag) | 1.81 (zigzag) |
| | | 116 (armchair) | 15.0 (armchair) | 14.3 (armchair) | 1.03 (armchair) |
| | 600K | 100 (zigzag) | 11.9 (zigzag) | 13.1 (zigzag) | 0.81 (zigzag) |
| | | 82 (armchair) | 9.7 (armchair) | 15.1 (armchair) | 0.84 (armchair) |
| 4/27, 1K | | 180 (zigzag) | 22.2 (zigzag) | 12.5 (zigzag) | 1.38 (zigzag) |
| | | 192 (armchair) | 24.8 (armchair) | 13.3 (armchair) | 1.55 (armchair) |
| | 300K | 95 (zigzag) | 14.0 (zigzag) | 18.0 (zigzag) | 1.36 (zigzag) |
| | | 96 (armchair) | 13.1 (armchair) | 15.0 (armchair) | 1.03 (armchair) |
| | 600K | 87 (zigzag) | 11.0 (zigzag) | 15.1 (zigzag) | 0.96 (zigzag) |
| | | 88 (armchair) | 9.5 (armchair) | 15.6 (armchair) | 0.95 (armchair) |

a

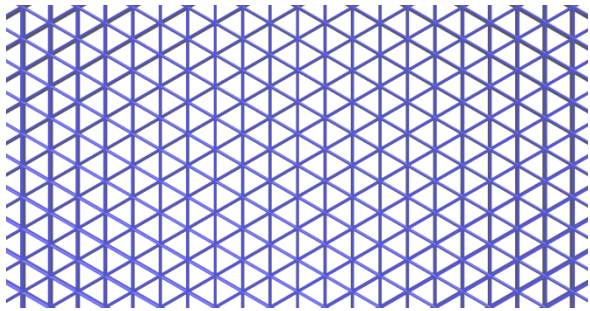
a) $\eta=0$, triangular boron sheet

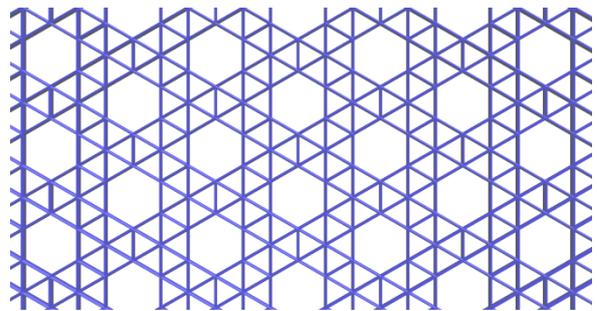
b) $\eta=1/9$ ($\alpha$ - boron sheet)

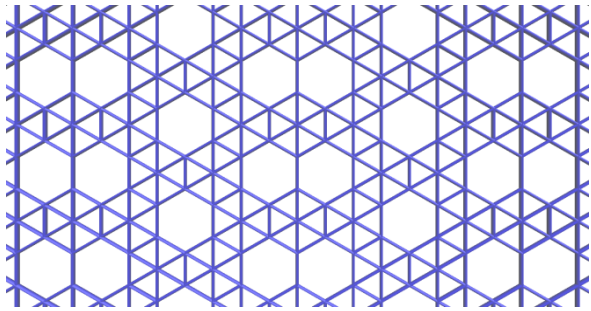
c) $\eta=1/8$ (type a)

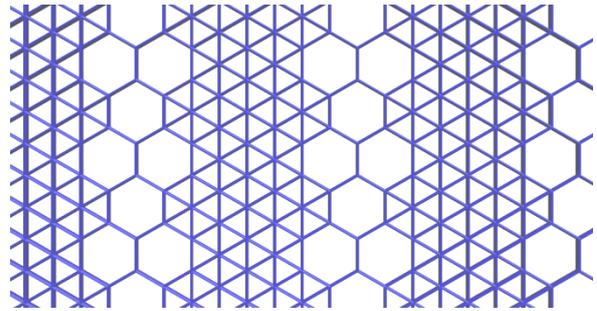
d) $\eta=1/8$ (type b)

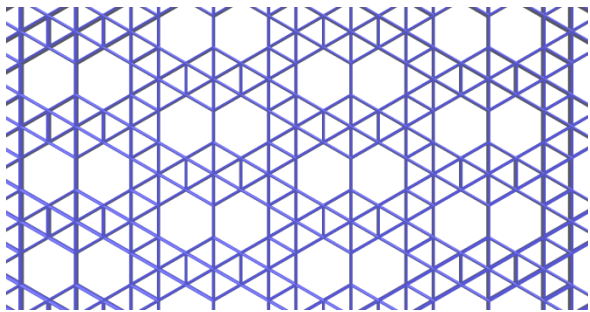
e) $\eta=2/15$

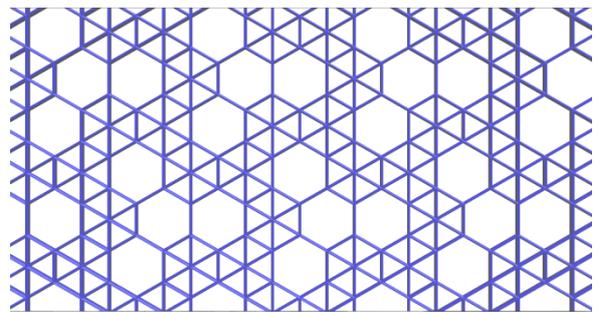
f) $\eta=4/27$

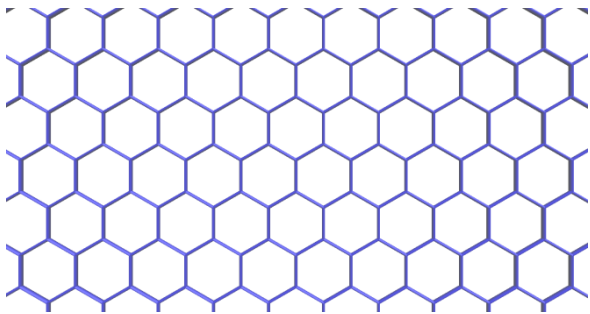
g) $\eta=1/3$, hexagonal boron sheet

Fig. 1 Schematic illustration of various boron sheets. Zigzag and armchair directions are horizontal and vertical, respectively. $\eta$ is the ratio between the number of hexagon holes and the number of atoms in the original triangular sheet.

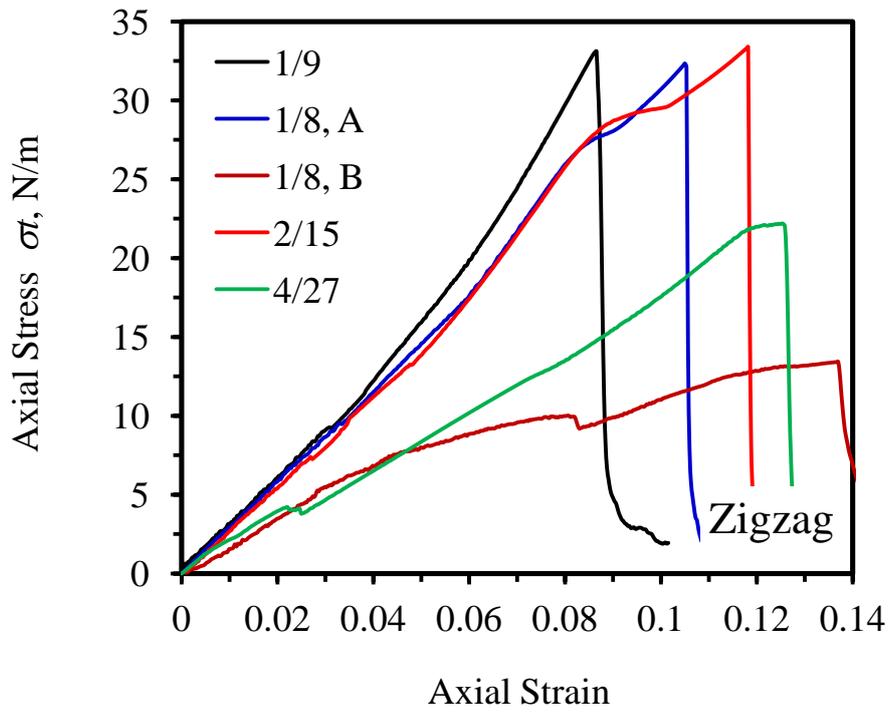

a)

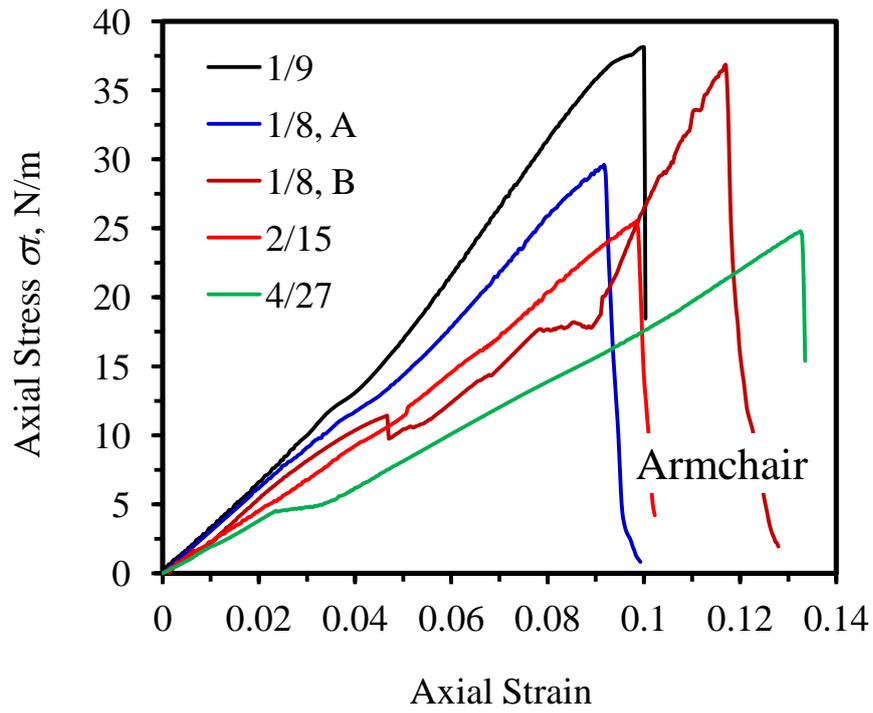

b)

Fig. 2 Evolution versus axial stress versus axial strain at 1 K under uniaxial tension in: a) zigzag; and b) armchair direction.

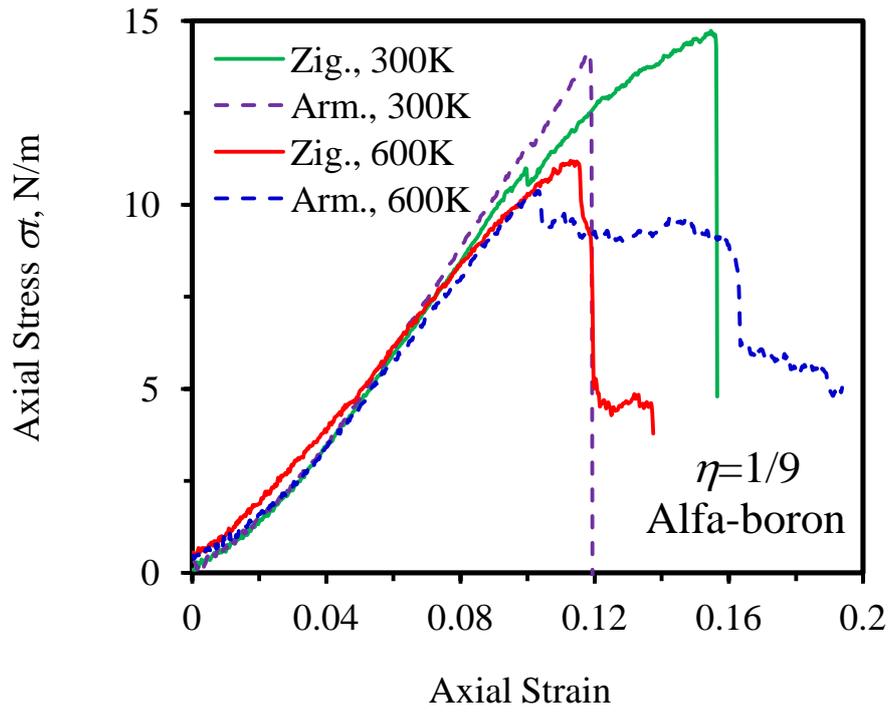

a)

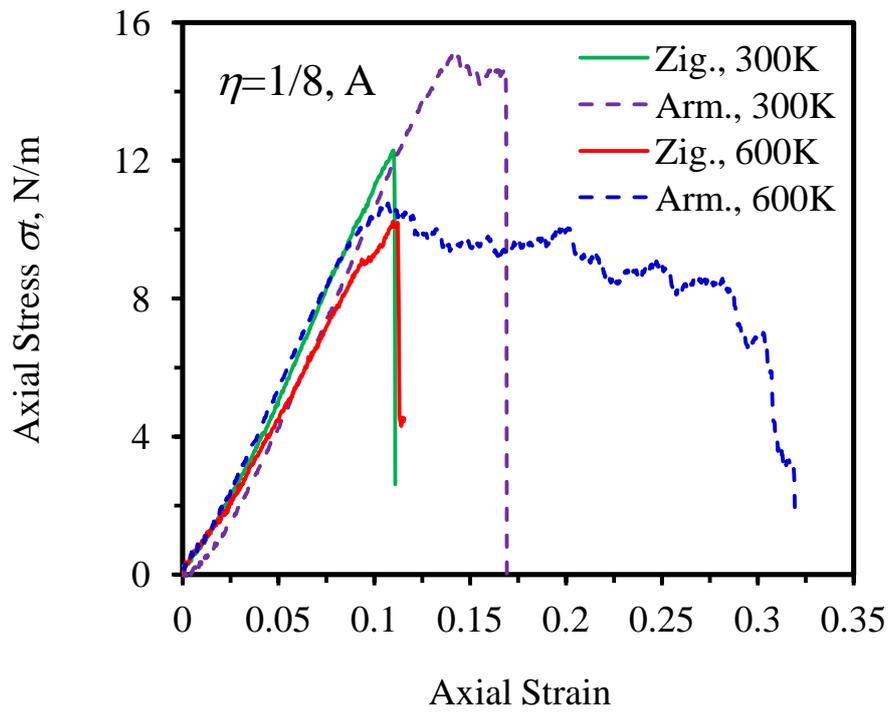

b)

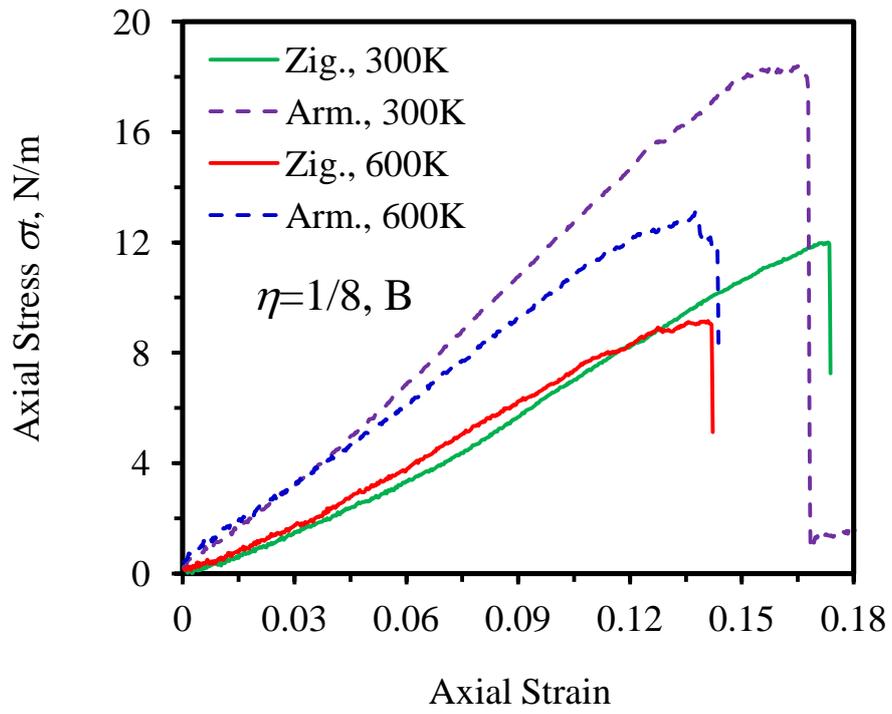

c)

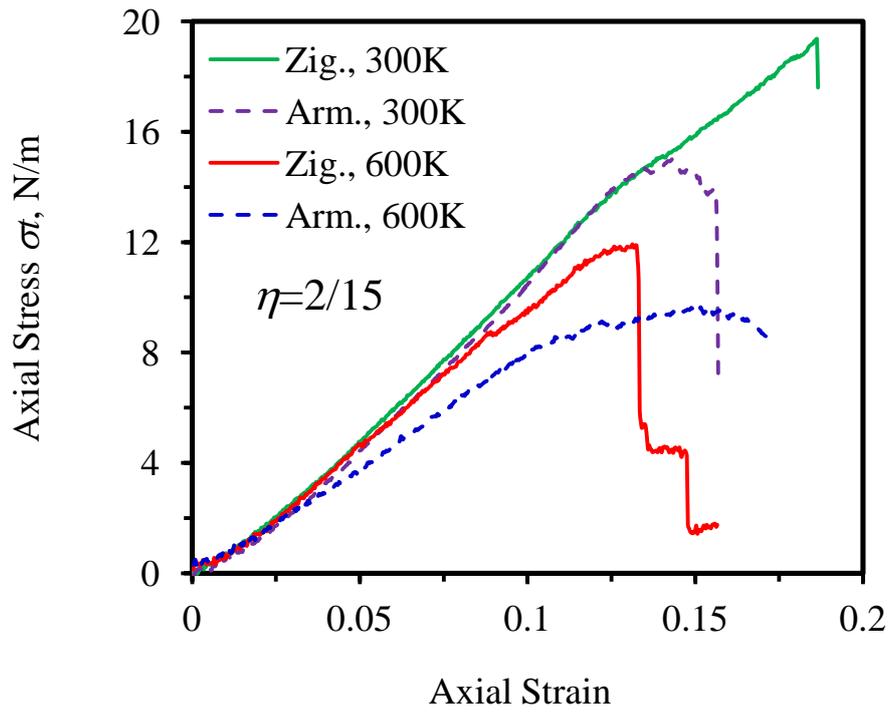

d)

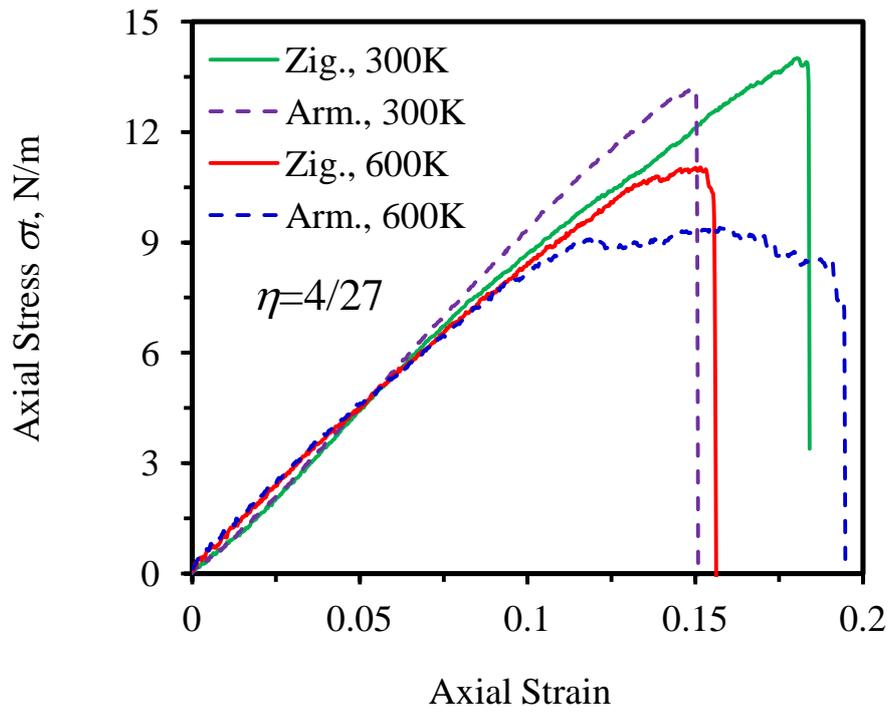

e)

Fig. 3 Evolution of axial stress versus axial strain at 300 K and 600K of 5 types of boron sheets under uniaxial tension: a) $\eta=1/9$; b) $\eta=1/8$, type A; c) $\eta=1/8$, type B; d) $\eta=2/15$; e) $\eta=4/27$.

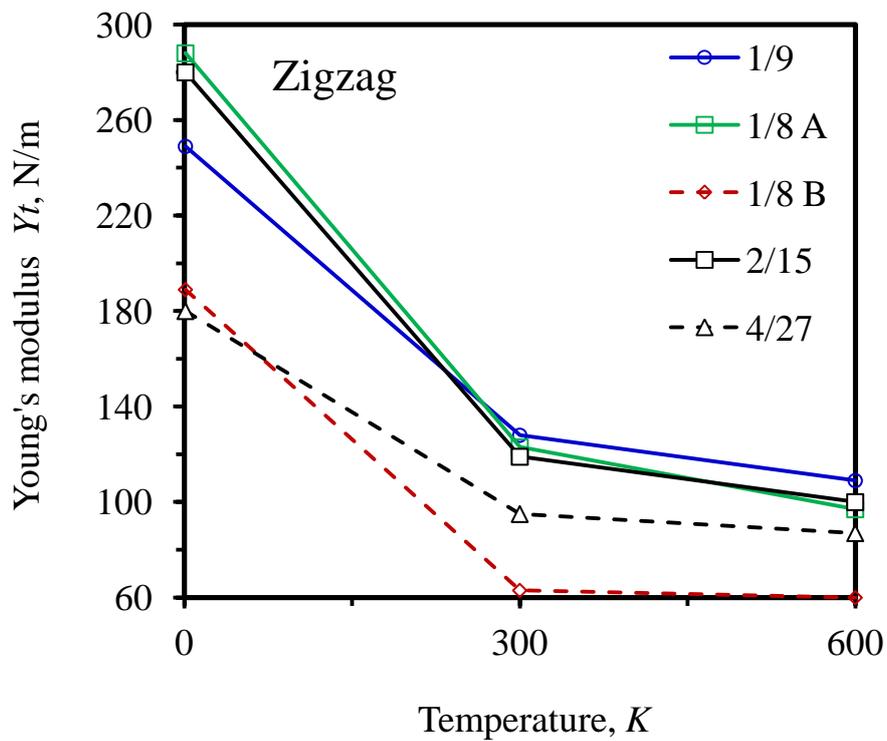

a)

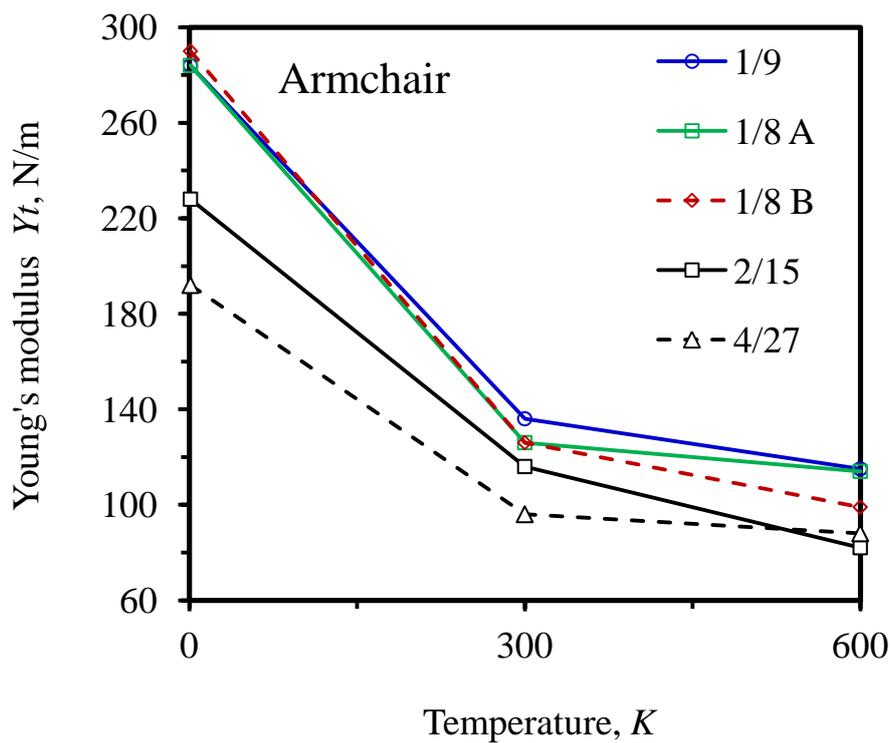

b)

Fig. 4 Effects of temperature on Young's modulus of boron sheets under uniaxial tension in the: a) zigzag direction; and b) armchair direction.

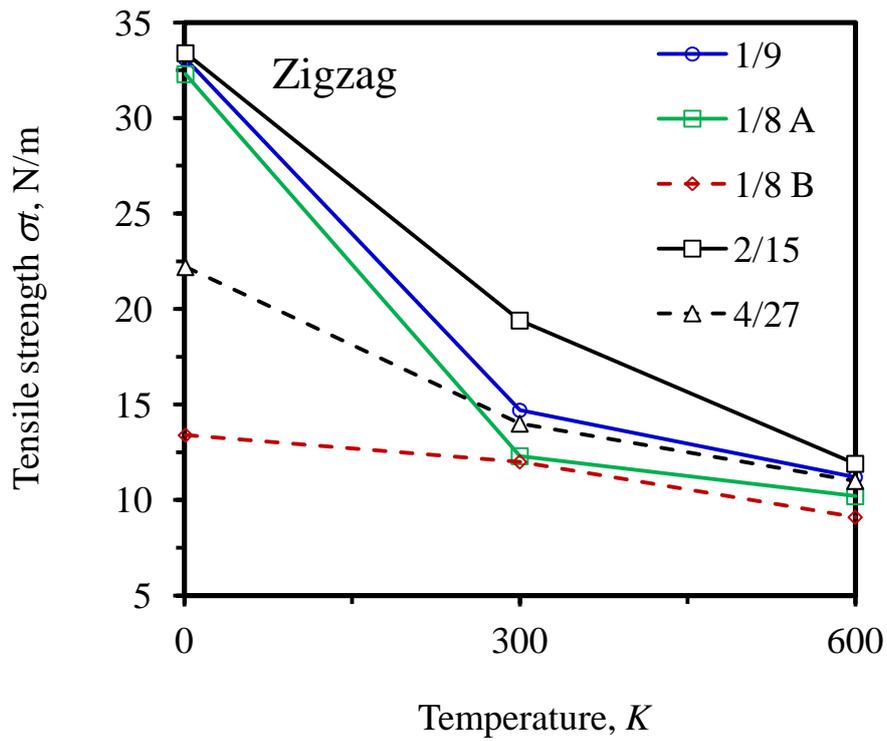

a)

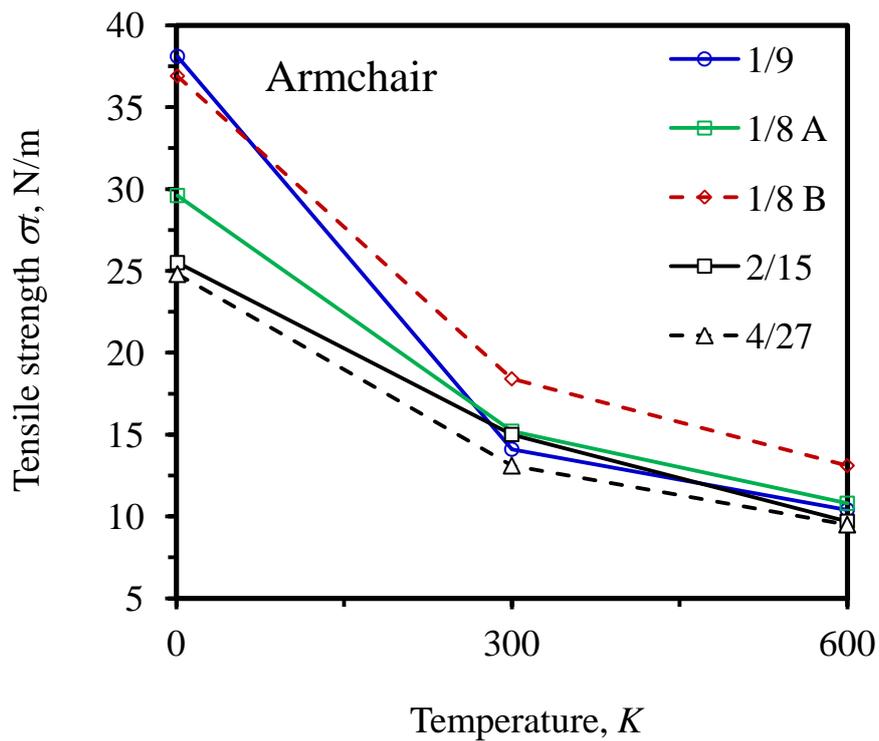

b)

Fig. 5 Effects of temperature on tensile strength of boron sheets under uniaxial tension in the: (**top**) zigzag direction; and (**bottom**) armchair direction.

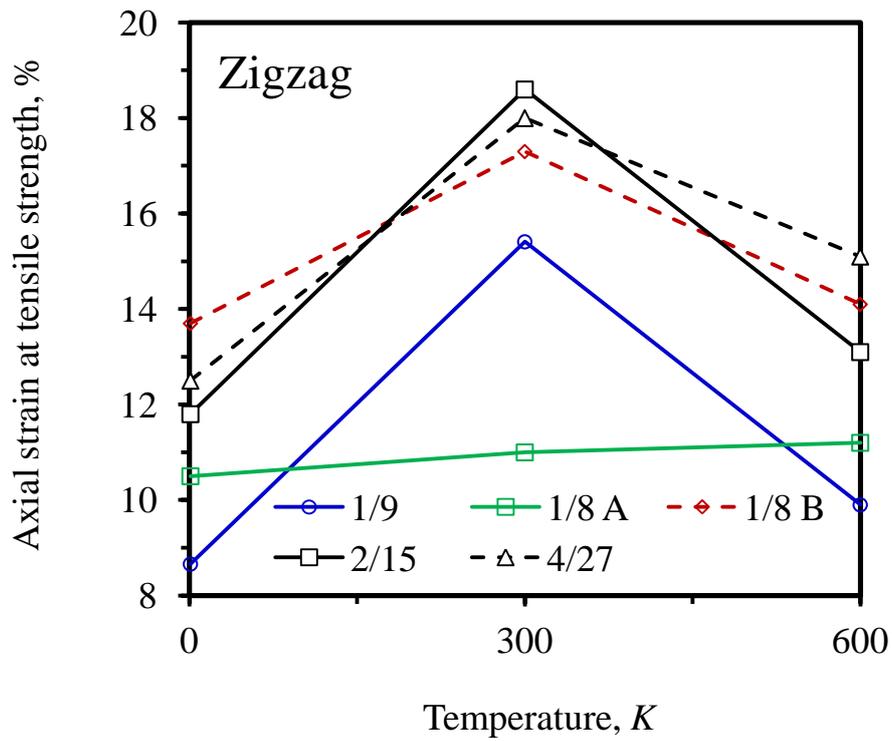

a)

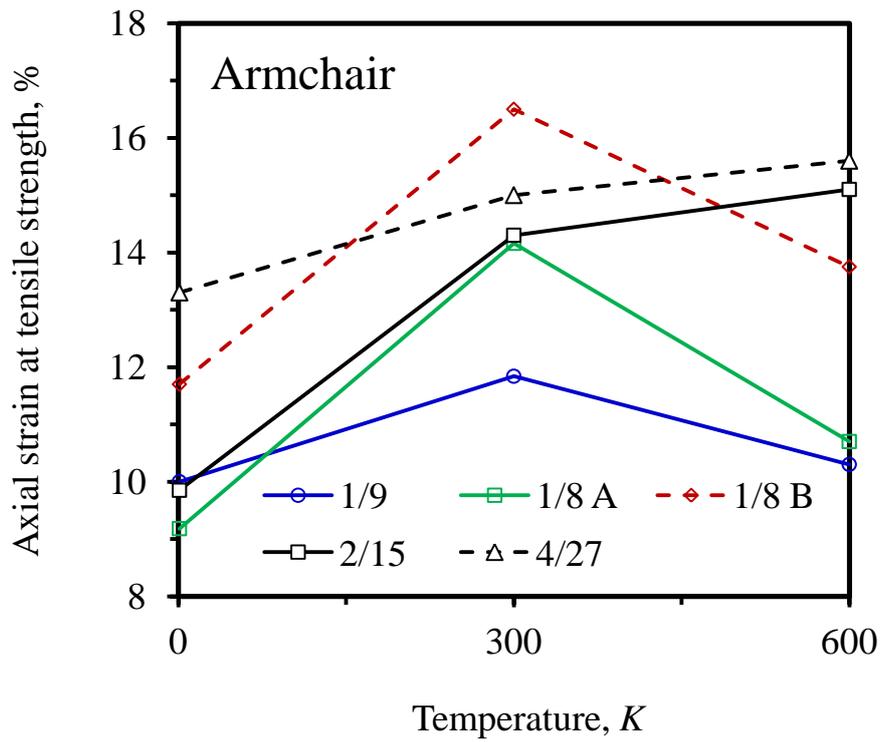

b)

Fig. 6 Effects of temperature on axial strain at tensile strength of boron sheets under uniaxial tension in the: a) zigzag direction; and b) armchair direction.

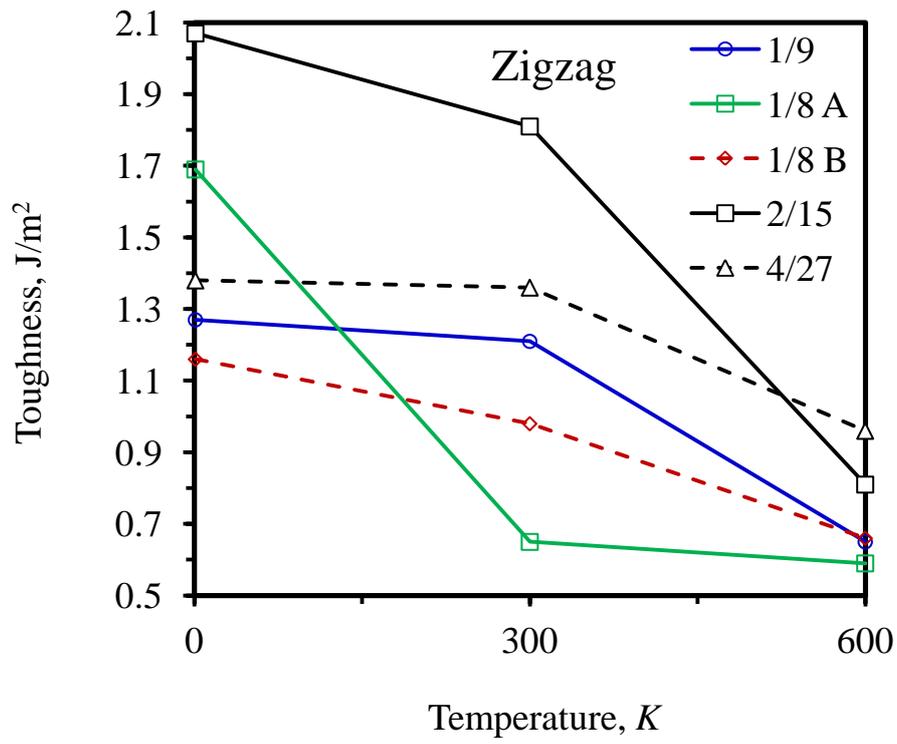

a)

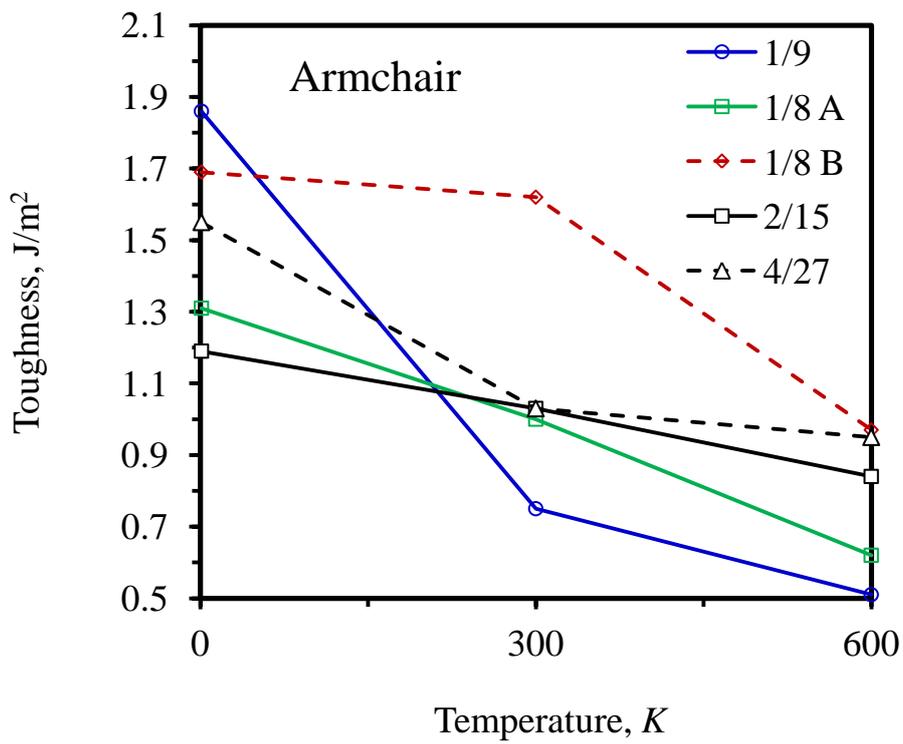

b)

Fig. 7 Effects of temperature on the toughness of boron sheets under uniaxial tension in the: a) zigzag direction; and b) armchair direction.

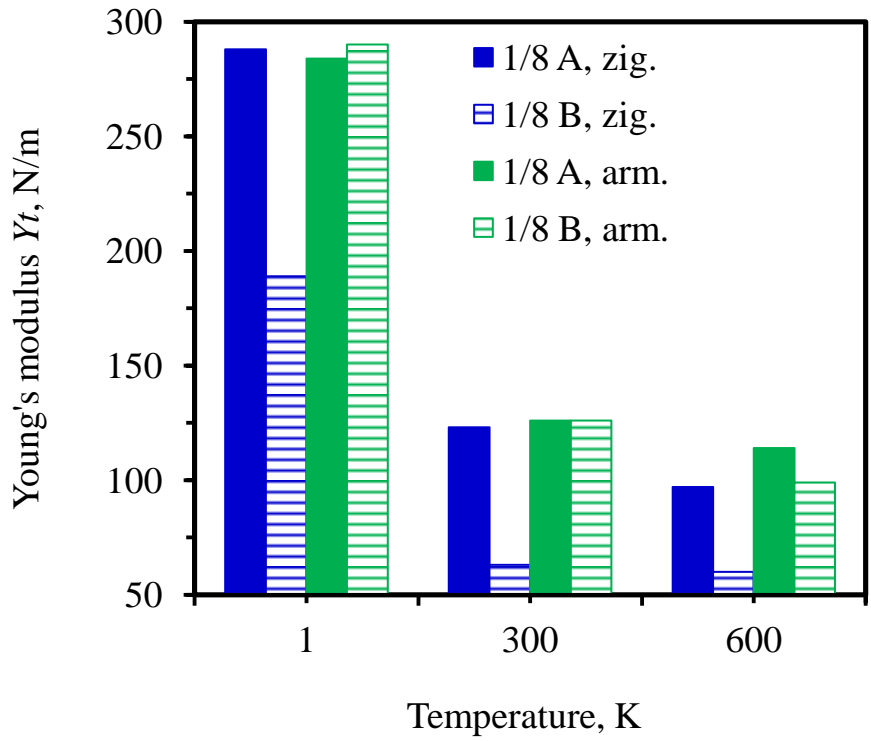

a)

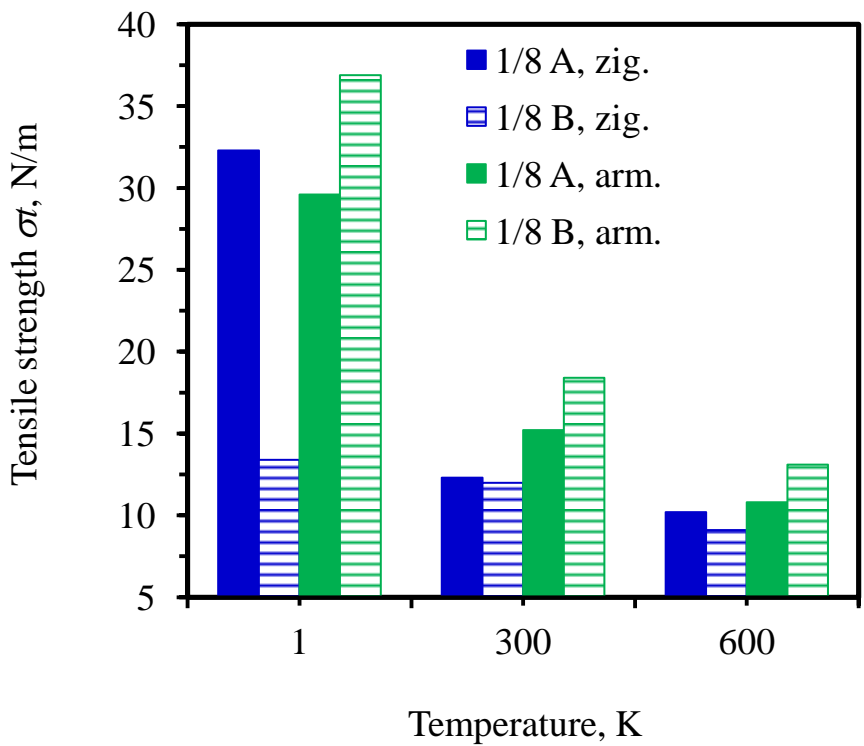

b)

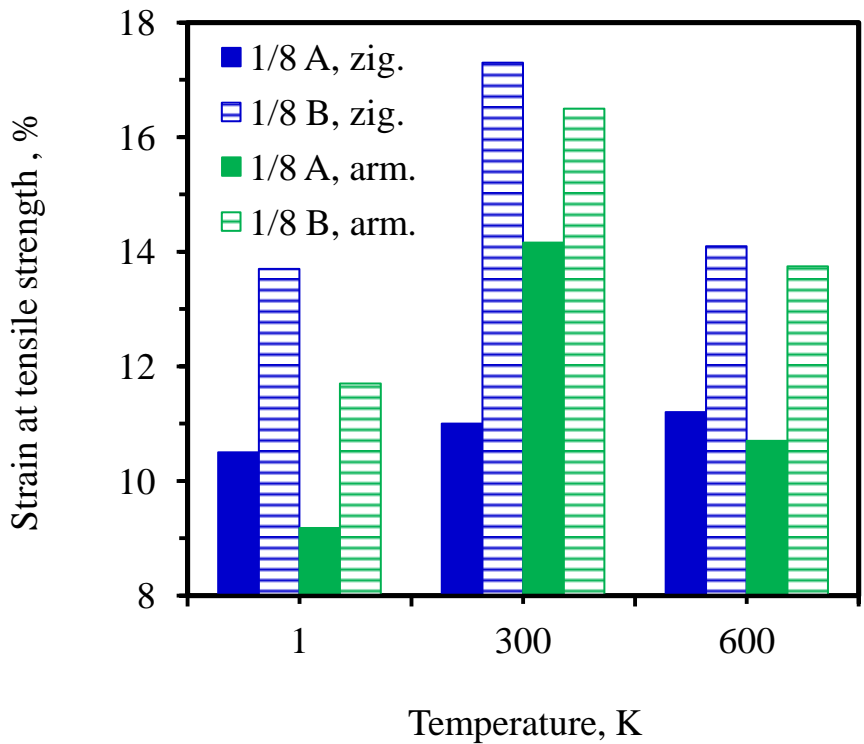

c)

Fig. 8 Comparison of mechanical properties of 2 boron sheets with $\eta=1/8$ (type A) and $\eta=1/8$ (type B): a) Young's modulus; b) tensile strength; and c) strain at tensile strength.

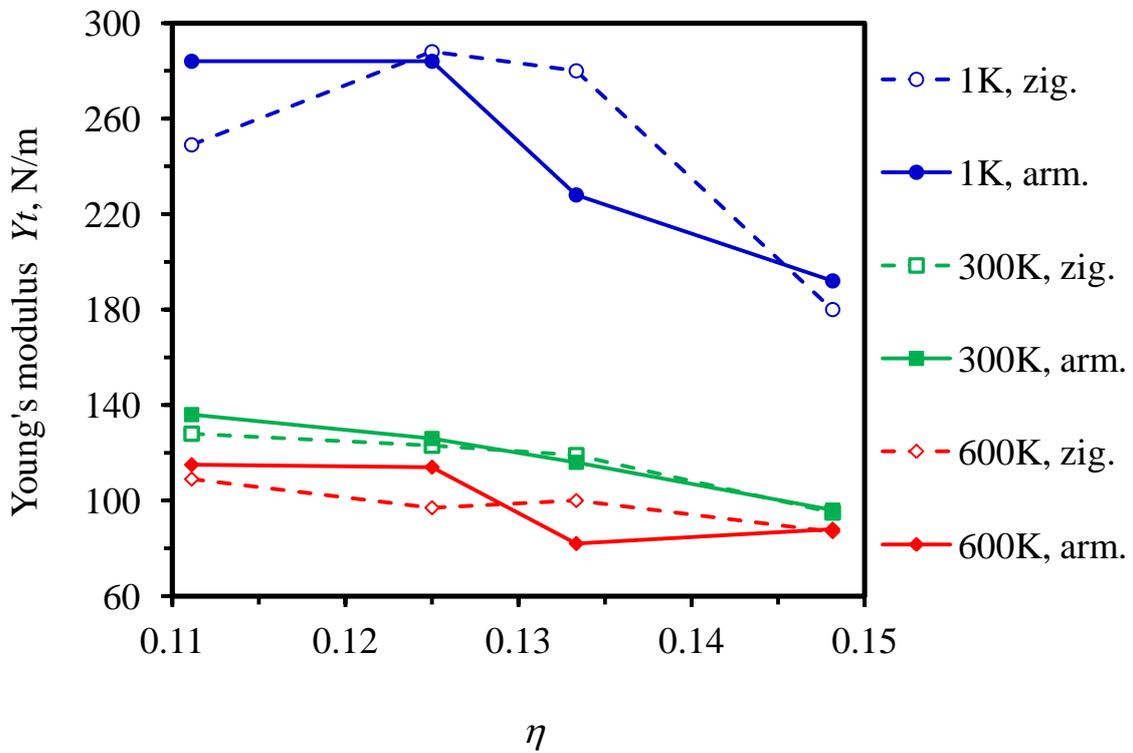

a)

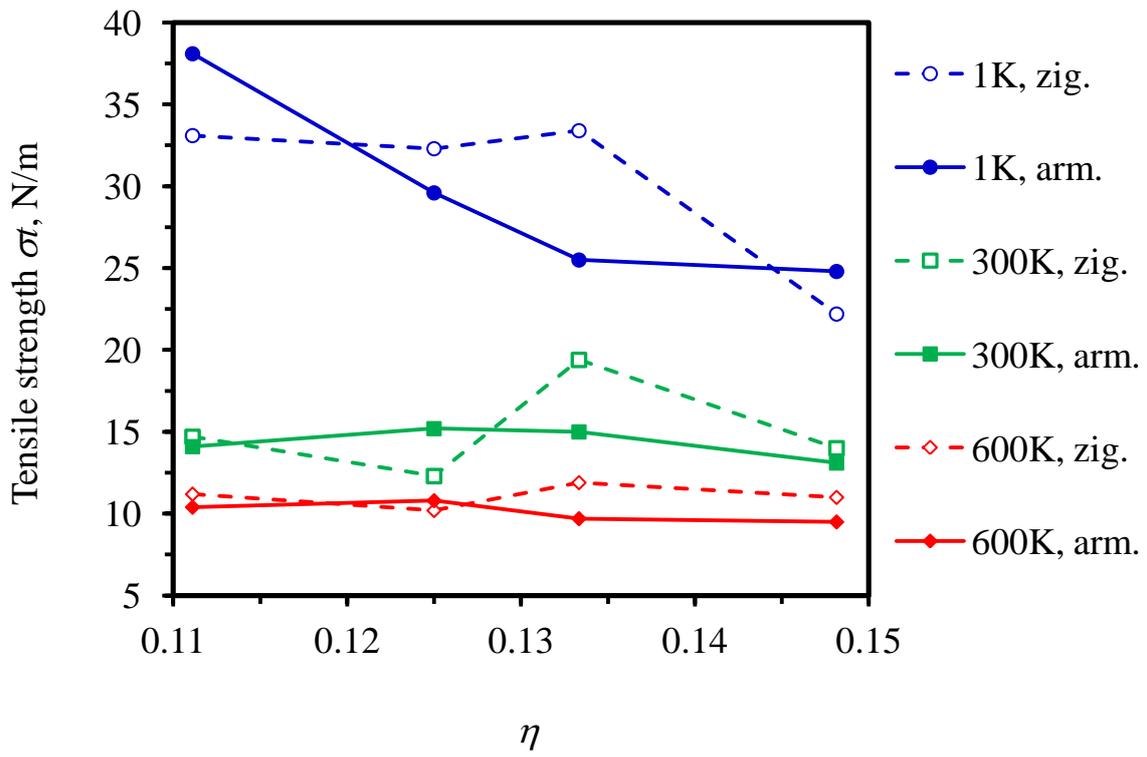

b)

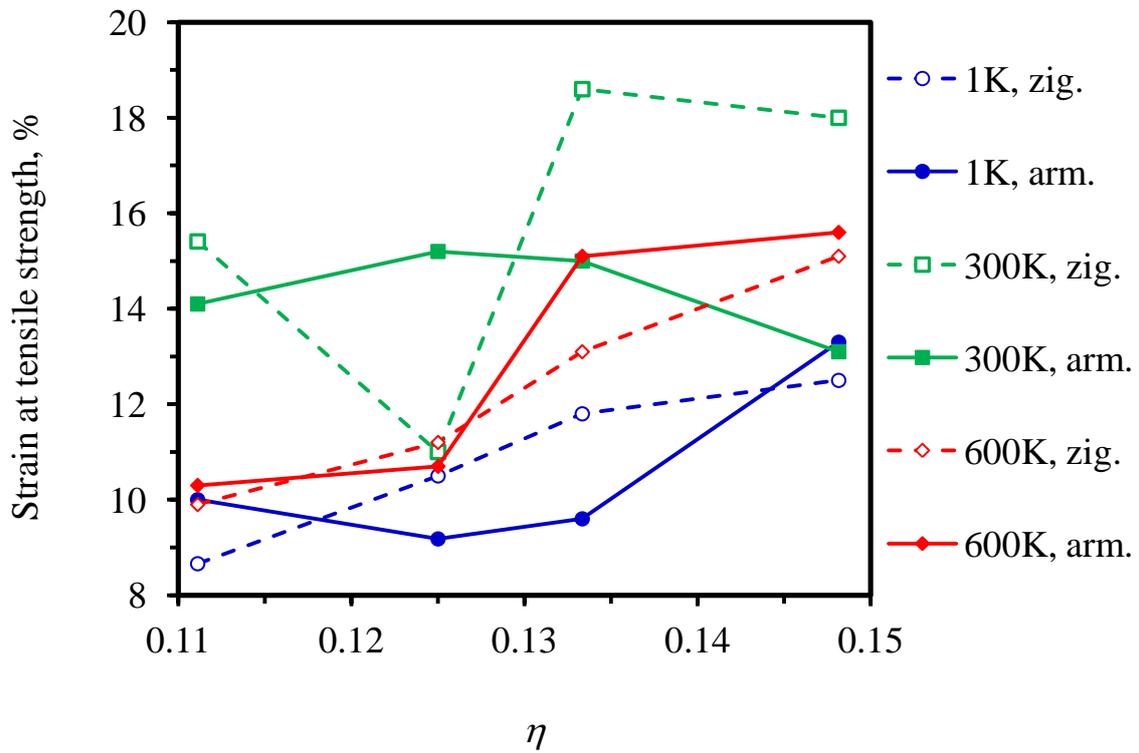

c)

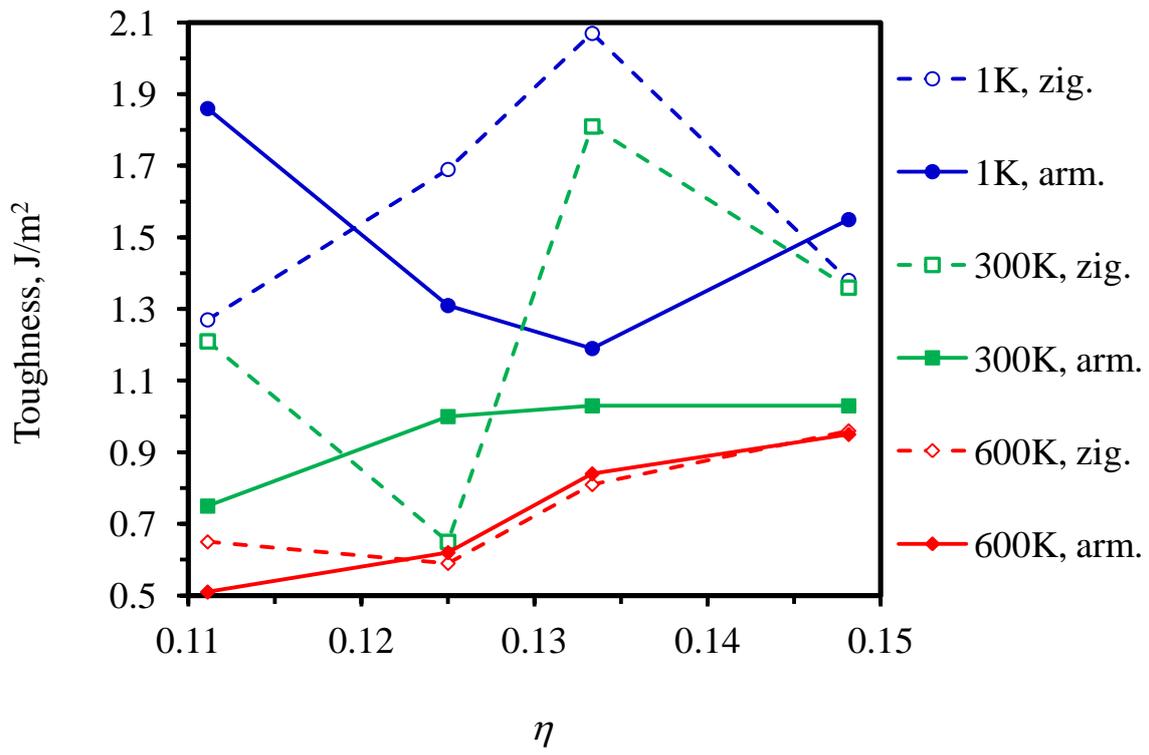

d)

Fig. 9 Effects of vacancy ratio $\eta$ on: a) Young's modulus; b) tensile strength; c) strain tensile strength; and d) toughness.

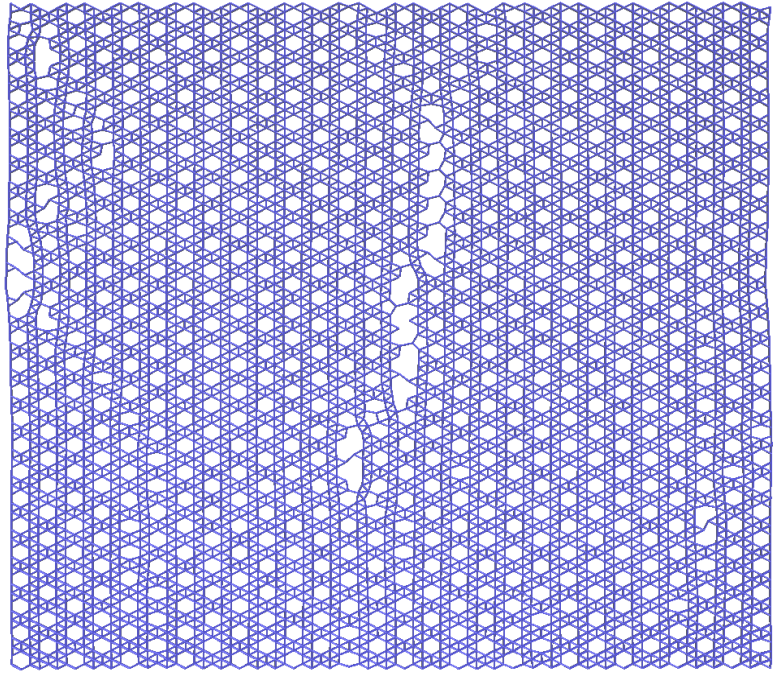

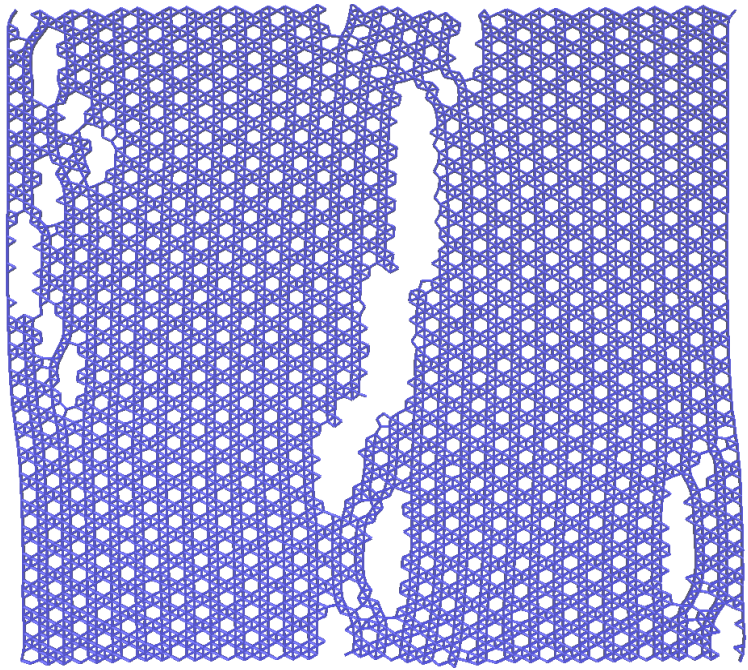

Fig. 10 Snapshots of $\alpha$-boron sheet ($\eta=1/9$) under uniaxial tension in the zigzag direction at 1 K: a) at axial tensile strain $\varepsilon=8.75\%$; and b) at $\varepsilon=9.0625\%$. The tensile strain at maximum axial stress is about 8.66%.

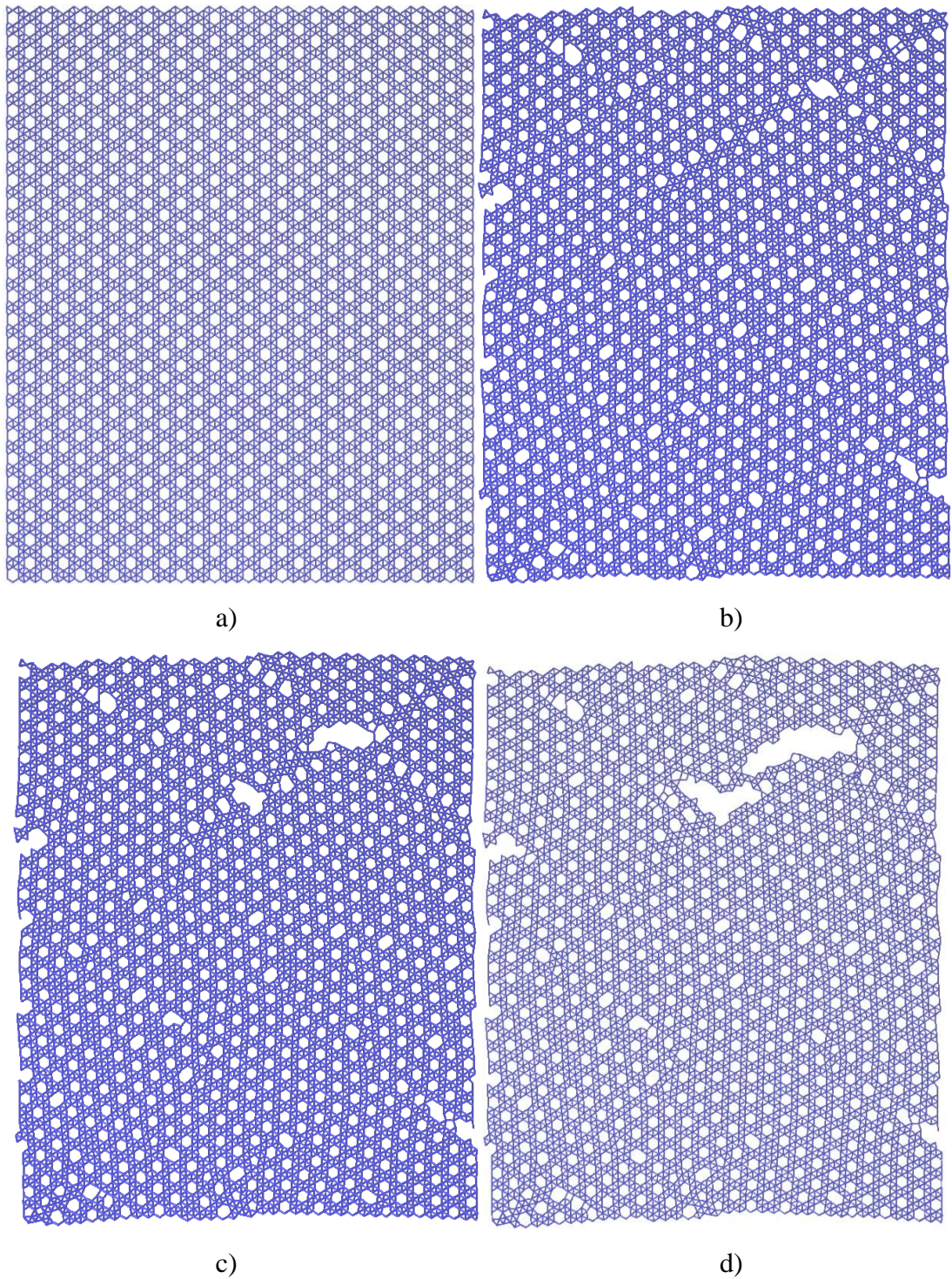

Fig. 11 Snapshots of α-boron sheet (η=1/9) under uniaxial tension in the armchair direction at 1 K: a) no evidence of failure at axial tensile strain at $\varepsilon$=10.0% at tensile strength; b) at $\varepsilon$=10.3125%; c) at $\varepsilon$= 10.9375%; and d) at $\varepsilon$=11.25%.

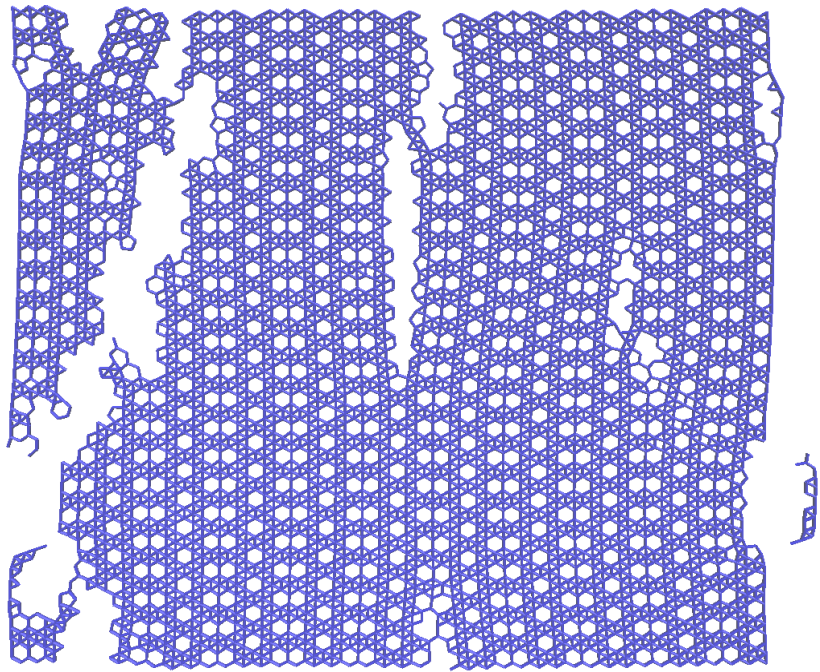

a)

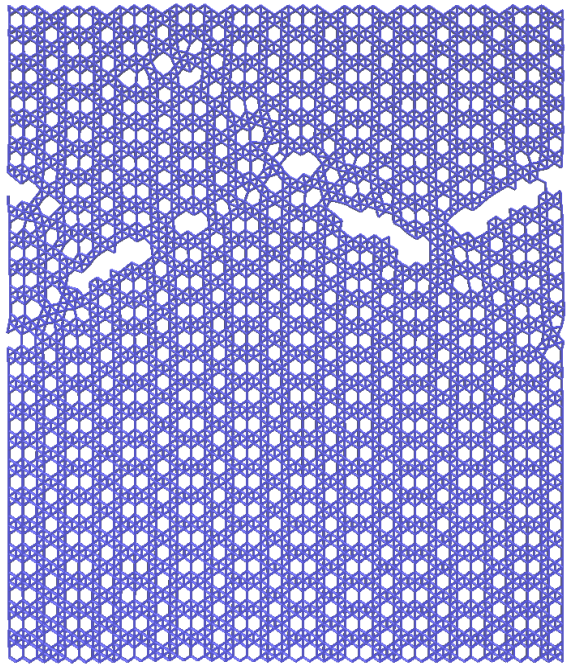

b)

Fig. 12 Snapshots of sheet with $\eta=1/8$ (type A) at 1 K under uniaxial tension: a) in the zigzag at axial tensile strain $\varepsilon=10.69$ %; b) in the armchair direction at $\varepsilon=9.3$ %. The tensile strain at maximum axial stress is about 10.5 % and 9.18% in the zigzag and armchair direction, respectively.

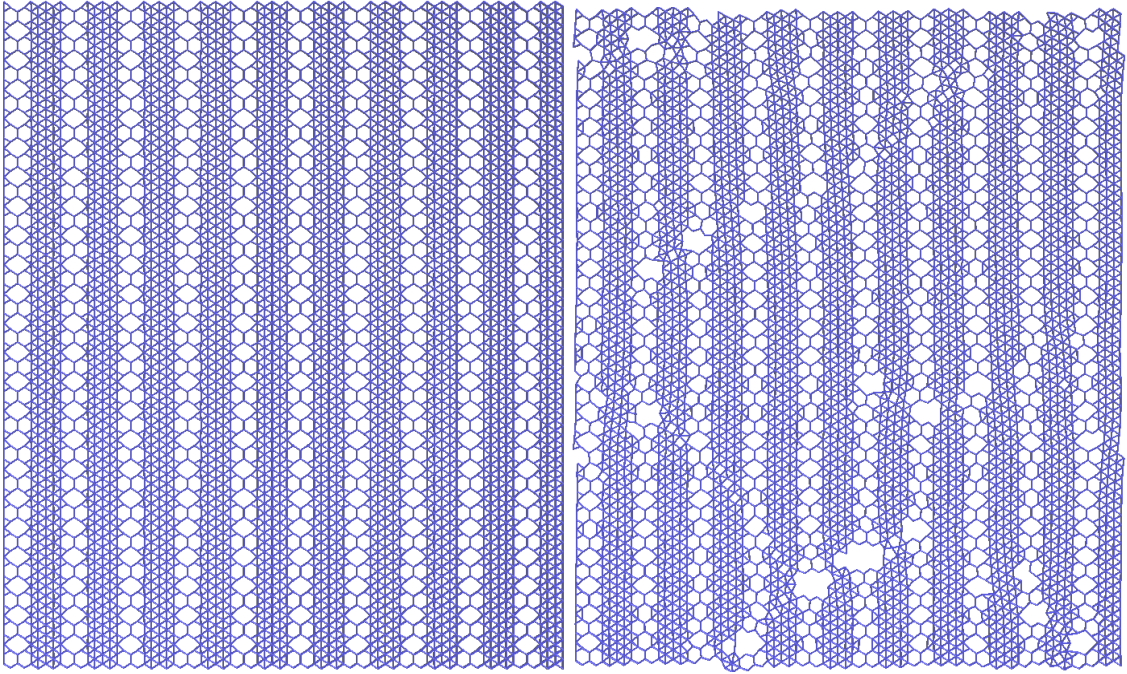

Fig. 13 Snapshots of sheet with $\eta=1/8$ (type B) under uniaxial tension in the armchair direction at 1 K: (**left**) no evidence of failure at axial tensile strain $\varepsilon=11.5\%$; (**right**) failure at $\varepsilon=11.9\%$. The tensile strain at maximum axial stress is about 11.7%.

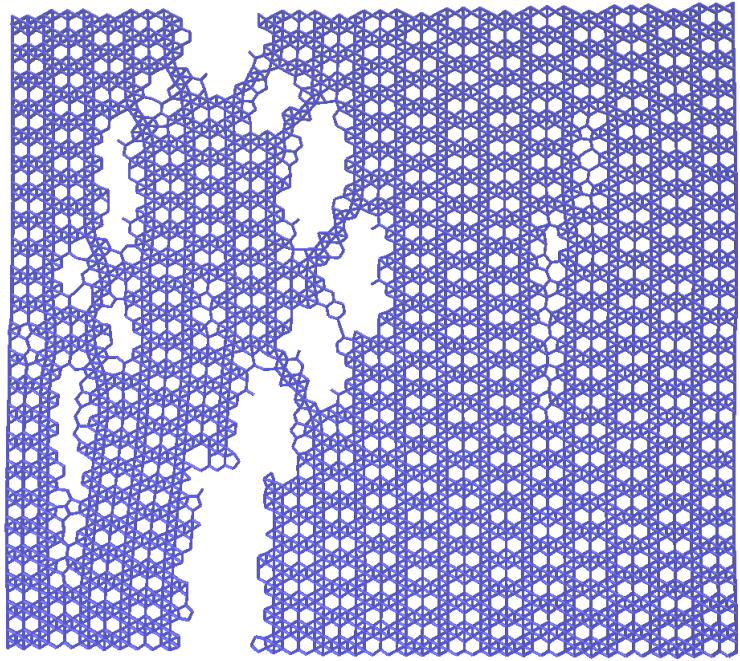

a)

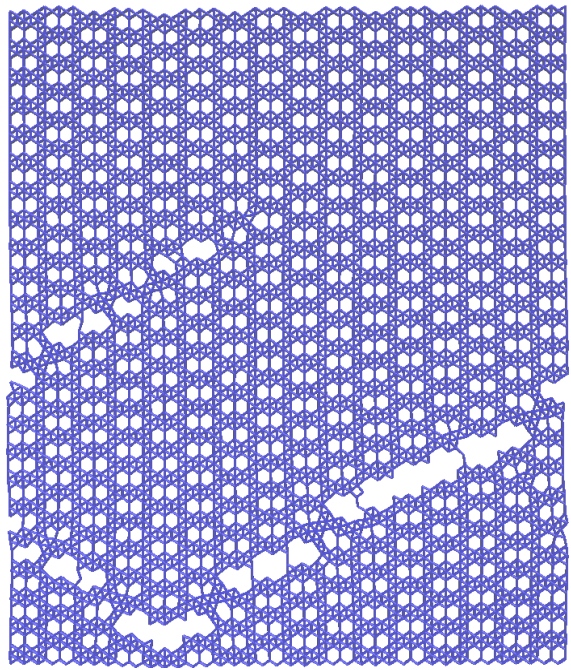

b)

Fig. 14 Snapshots of sheet with $\eta=2/15$ at 1 K under uniaxial tension: a) in the zigzag at axial tensile strain $\varepsilon=12.1$ %; b) in the armchair direction at $\varepsilon=10$ %. The tensile strain at maximum axial stress is about 11.82 % and 9.85 % in the zigzag and armchair direction, respectively.

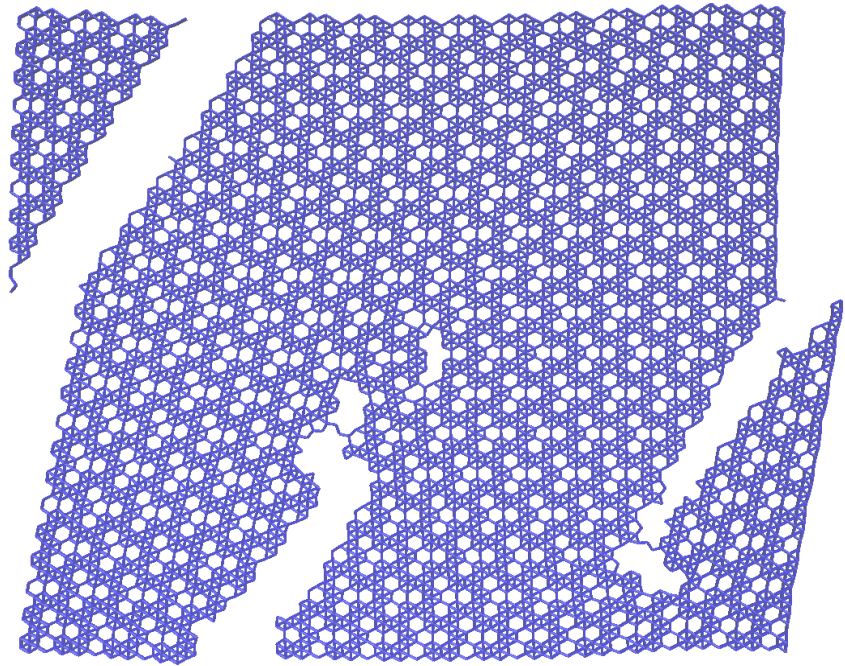

a)

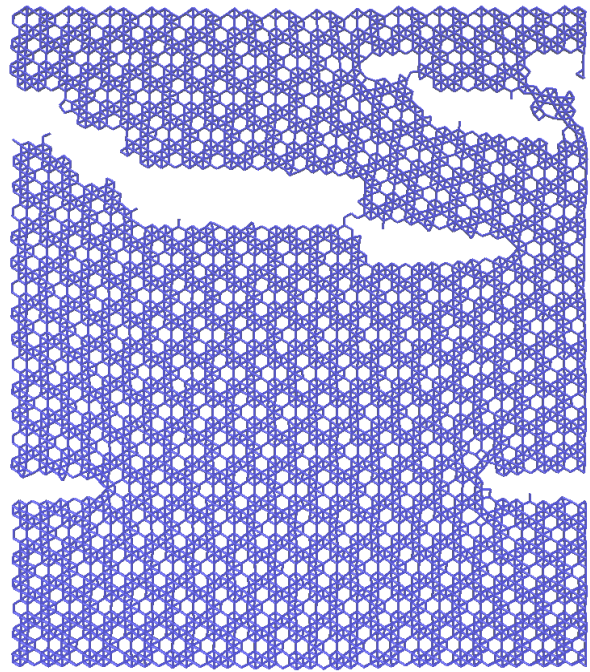

b)

Fig. 15 Snapshots of sheet with $\eta=4/27$ at 1 K under uniaxial tension: a) in the zigzag at axial tensile strain $\varepsilon=12.9$ %; b) in the armchair direction at $\varepsilon=13.4$ %. The tensile strain at maximum axial stress is about 12.5 % and 13.3 % in the zigzag and armchair direction, respectively.